      [1999/12/01 v1.4c Il Nuovo Cimento]
\documentclass{cimento}

\usepackage{graphicx}
\usepackage{dcolumn}
\usepackage{amssymb}
\usepackage{amsmath}
\usepackage{bm}
\title{Equatorial Circular Geodesics in the Hartle-Thorne Spacetime}
\author{D.~Bini\from{ins:z}\from{ins:p}\from{ins:q}\ETC,
K.~Boshkayev\from{ins:p}\from{ins:y},
R.~Ruffini\from{ins:p}\from{ins:y}\from{ins:x}\atque
I.~Siutsou\from{ins:p}\from{ins:y}}
\instlist{\inst{ins:z} Istituto per le Applicazioni del Calcolo "M. Picone," CNR, I-00185 Rome, Italy
  \inst{ins:p} Dipartimento di Fisica and ICRA, Universit\`a di Roma Sapienza,\\ P.le Aldo Moro 5, I-00185 Roma, Italy
  \inst{ins:q} INFN, Sezione di Firenze, I–00185 Sesto Fiorentino (FI), Italy
  \inst{ins:y} ICRANet, Piazzale della Repubblica 10, I-65122 Pescara, Italy 
  \inst{ins:x} Universite de Nice Sophia Antipolis, CEDEX 2, Grand Chateau Parc Valrose, Nice, France.}
\PACSes{\PACSit{}{04.20.-q 	Classical general relativity. 95.30.Sf 	Relativity and gravitation}}

\begin{document}
\maketitle
\begin{abstract}
We investigate the influence of the quadrupole moment of a rotating source on the motion of a test particle in the strong field regime. For this purpose the Hartle-Thorne metric, that is an approximate solution of vacuum Einstein field equations that describes the exterior of any slowly rotating, stationary and axially symmetric body, is used. The metric is given with accuracy up to the second order terms in the body's angular momentum, and first order terms in its quadrupole moment. We give, with the same accuracy, analytic equations for equatorial circular geodesics in the Hartle-Thorne spacetime and integrate them numerically.
\end{abstract}
\section{Introduction}
Astrophysical objects in general are characterized by a non-spherically symmetric distribution of mass. In many cases, like ordinary planets and satellites, it is possible to neglect the deviations from spherical symmetry: it seems instead reasonable to expect that deviations should be taken into account in case of strong gravitational fields. The metric describing the exterior field of a slowly rotating slightly deformed object was found by Hartle and Thorne \cite{H,HT}. However in this work we use the form of the metric presented in \cite{Bini2009}. In geometrical units it is given by
\begin{eqnarray}\label{ht1}
ds^2&=&-\left(1-\frac{2{ M }}{r}\right)\left[1+2k_1P_2(\cos\theta)+2\left(1-\frac{2{M}}{r}\right)^{-1}\frac{J^{2}}{r^{4}}(2\cos^2\theta-1)\right]dt^2\\\nonumber
&+&\left(1-\frac{2{M}}{r}\right)^{-1}\left[1-2k_2P_2(\cos\theta)-2\left(1-\frac{2{M}}{r}\right)^{-1}\frac{J^{2}}{r^4}\right]dr^2\\\nonumber
&+&r^2[1-2k_3P_2(\cos\theta)](d\theta^2+\sin^2\theta d\phi^2)-\frac{4J}{r}\sin^2\theta dt d\phi\,
\end{eqnarray}
where
\begin{eqnarray}\label{ht2}
k_1&=&\frac{J^{2}}{{M}r^3}\left(1+\frac{{M}}{r}\right)-\frac{5}{8}\frac{Q-J^{2}/{M}}{{M}^3}Q_2^2\left(\frac{r}{{M}}-1\right), \quad k_2=k_1-\frac{6J^{2}}{r^4}, \nonumber\\
k_3&=&k_1+\frac{J^{2}}{r^4}-\frac{5}{4}\frac{Q-J^{2}/{M}}{{M}^2r}\left(1-\frac{2{M}}{r}\right)^{-1/2}Q_2^1\left(\frac{r}{M}-1\right),\ P_{2}(x)=\frac{1}{2}(3x^{2}-1),\nonumber\\
Q_{2}^{1}(x)&=&(x^{2}-1)^{1/2}\left[\frac{3x}{2}\ln\frac{x+1}{x-1}-\frac{3x^{2}-2}{x^{2}-1}\right],\ Q_{2}^{2}(x)=(x^{2}-1)\left[\frac{3}{2}\ln\frac{x+1}{x-1}-\frac{3x^{3}-5x}{(x^{2}-1)^2}\right].\nonumber
\end{eqnarray}
Here $P_{2}(x)$ is Legendre polynomials of the first kind, $Q_l^m$ are the associated Legendre polynomials of the second kind and the constants ${M}$, ${J}$ and ${Q}$ are the total mass, angular momentum and  quadrupole parameter of a rotating star respectively\footnote{We note here that the quadrupole parameter $Q$ is related to the mass quadrupole moment defined by Hartle and Thorne \cite{HT} through $Q=2J^2/M-Q_{HT}$.}. The approximate Kerr metric \cite{K} in the Boyer-Lindquist coordinates $(t,\ R,\ \Theta,\ \phi)$ up to second order terms in the rotation parameter $a$ can be obtained from (\ref{ht1}) by setting
\begin{equation}\label{tr1}
J=-Ma,\quad Q={J}^2/{M},
\end{equation}
and making a coordinate transformation given by
\begin{eqnarray}\label{tr2}
r&=&R+\frac{a^2}{2R}\left[\left(1+\frac{2M}{R}\right)\left(1-\frac{M}{R}\right)-\cos^2\Theta\left(1-\frac{2M}{R}\right)\left(1+\frac{3M}{R}\right)\right], \\
\theta&=&\Theta+\frac{a^2}{2R^2}\left(1+\frac{2M}{R}\right)\sin\Theta\cos\Theta. \nonumber\
\end{eqnarray}
\section{The Domain of validity of the Hartle-Thorne approximation}
Having on mind an application of the metric (\ref{ht1}) to the exterior of a compact object, we demand that the energy-momentum tensor, which follows from (\ref{ht1}), be much smaller than the corresponding tensor of the source object. A correct comparison of these tensors should be performed in terms of eigenvalues.
Consider a surface of the object which generates the metric under consideration. According to the Einstein equations $G_\alpha{}^\beta=8\pi T_\alpha{}^\beta$, where $\alpha,\ \beta=(t,\ r,\ \theta,\ \phi)$, the eigenvalues of the Einstein tensor inside the matter are equal to its density and pressure multiplied by $8\pi$ \cite{Synge1964}. Due to the inequality $\rho>p$ which holds for all known types of matter, the maximum of eigenvalues can be estimated as $8\pi\rho$, where $\rho$ represents the average density of the body:
\begin{equation}\label{crude}
|G_\alpha{}^\beta| \lesssim 8\pi\rho=\frac{8\pi M}{4\pi r^3/3}=\frac{6M}{r^3}.
\end{equation}
On the other hand, the first non-vanishing terms in the expansion of the Einstein tensor of the Hartle-Thorne metric in powers of $J$ and $Q$ are $G_0{}^4$ and $G_4{}^0$. Then the Einstein tensor has two purely imaginary eigenvalues different from zero $\lambda_{1,2}\neq0$ and two exactly zero eigenvalues $\lambda_{3,4}=0$. The first pair is diverging as $r\rightarrow2M$. Near this radius we have, for $\delta r=r-2M$ approaching $0$, the leading terms equal to
\begin{equation}\label{leading}
\lambda_{1,2}\rightarrow\pm i\frac{15 J Q (1-3 \cos^2\theta) \sin\theta}{32\sqrt2 M^{11/2} \delta r^{3/2}}.
\end{equation}
Finally, by comparing the absolute values of (\ref{crude}) and (\ref{leading}) for $r\rightarrow2M$, taking into account that $0\leq(1-3 \cos^2 \theta)^2 \sin^2\theta\leq1$, we obtain the following inequality, describing the domain of validity of the Hartle-Thorne metric around the gravitating body
\begin{equation}
\delta r^3\gg\frac{25 J^2 Q^2}{128 M^7}.
\end{equation}
If we take the extreme values of the parameters for neutron stars such as $J\simeq M^2$, $Q\simeq10^{-2}M^3$ we obtain $\delta r\gg 3\times10^{-2}M$, that is certainly true for the exterior of neutron stars while their radii are more than $2.5M$ \cite{Haensel}, i.\,e. $\delta r > 0.5M$.
\section{Equations For the Equatorial Circular Geodesics}
\subsection{The Orbital Angular Velocity} The 4-velocity $U$ of a test particle on a circular orbit can be parametrized by the constant angular velocity with respect to infinity $\zeta$ 
\begin{equation}
U=\Gamma[\partial_t+\zeta\partial_{\phi}],
\end{equation}
where $\Gamma$ is a normalization factor which assures that $U^{\alpha}U_{\alpha}=-1$. From the normalization and the geodesics conditions we obtain the following expressions for $\Gamma$ and $\zeta=U^{\phi}/U^{t}$
\begin{eqnarray}\label{eight}
g_{tt}+2\zeta g_{t\phi}+\zeta^2 g_{\phi\phi}&=&-1/\Gamma^2,\quad g_{tt,r}+2\zeta g_{t\phi,r}+\zeta^2 g_{\phi\phi,r}=0,
\end{eqnarray}
where $g_{\alpha\beta,r}=\partial g_{\alpha\beta}/\partial r$. Hence, $\zeta$, the solution of (\ref{eight})$_2$, is given by
\begin{equation}
\zeta_{\pm}(u)=\pm\zeta_{0}(u)\left[1\mp j f_{1}(u)+j^2 f_{2}(u)+q f_{3}(u)\right]
\end{equation}
where $(+/-)$ stands for co-rotating/contra-rotating geodesics, $j=J/M^2$ and $q=Q/M^3$ are the dimensionless angular momentum and quadrupole parameter and $u=M/r$. The rest quantities are defined as follows
\begin{eqnarray}
\vspace{-1cm} 
\zeta_{0}(u)&=&\frac{u^{3/2}}{M},\quad f_{1}(u)=u^{3/2}, \nonumber \\
f_{2}(u)&=&\frac{48u^7-80u^6+4u^5+42u^4-40u^3-10u^2-15u+15}{16u^2(1-2u)}-f(u),\nonumber \\
f_{3}(u)&=&-\frac{5(6u^4-8u^3-2u^2-3u+3)}{16u^2(1-2u)}+f(u),\nonumber \\
f(u)&=&\frac{15(1-2u^3)}{32u^3}\ln\left(\frac{1}{1-2u}\right).\nonumber 
\end{eqnarray}
%
%
\begin{figure}*[t]
\hspace{-0.4cm} 
$\begin{array}{ccc}
\includegraphics[scale=0.325]{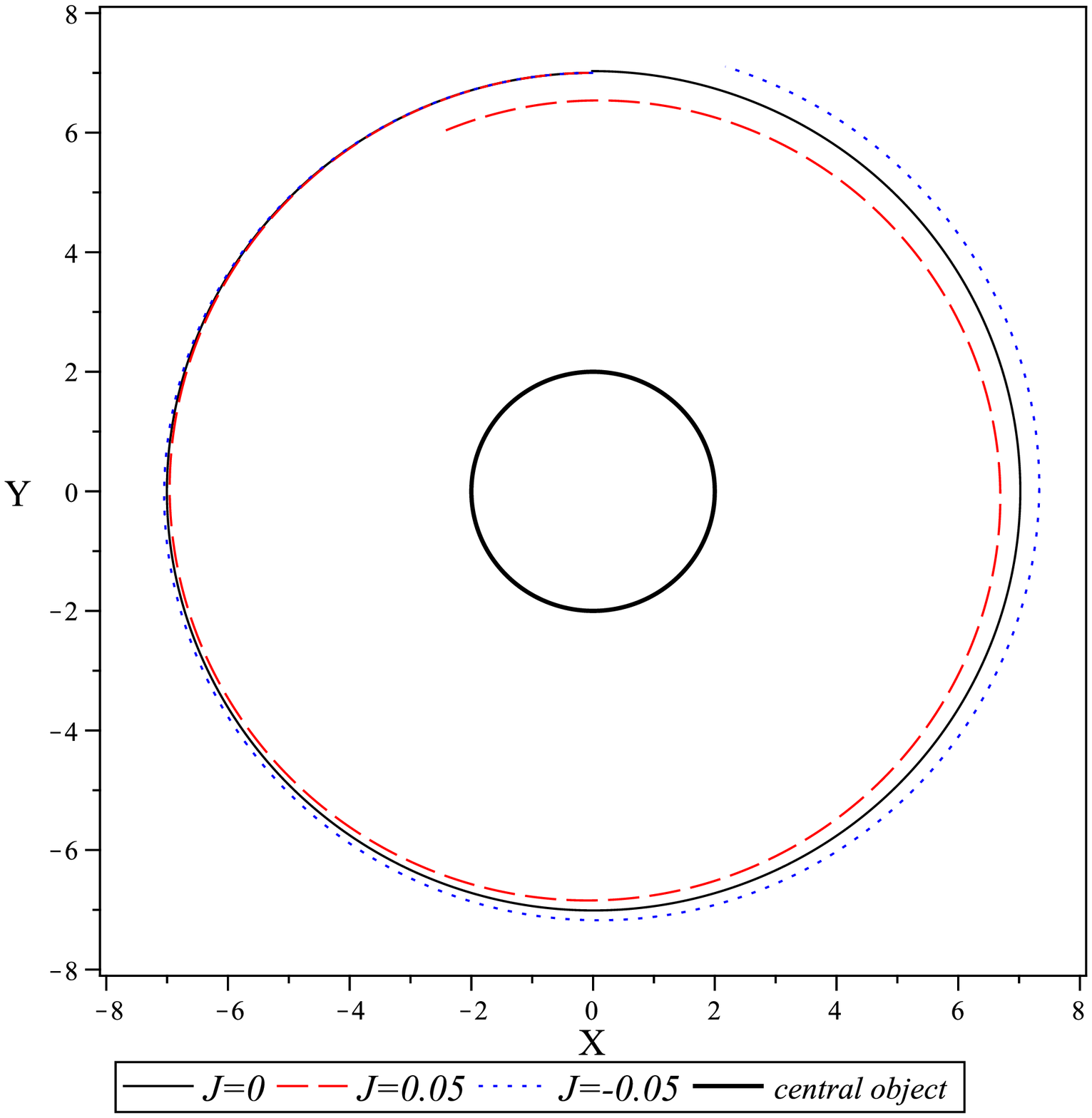} & \includegraphics[scale=0.325]{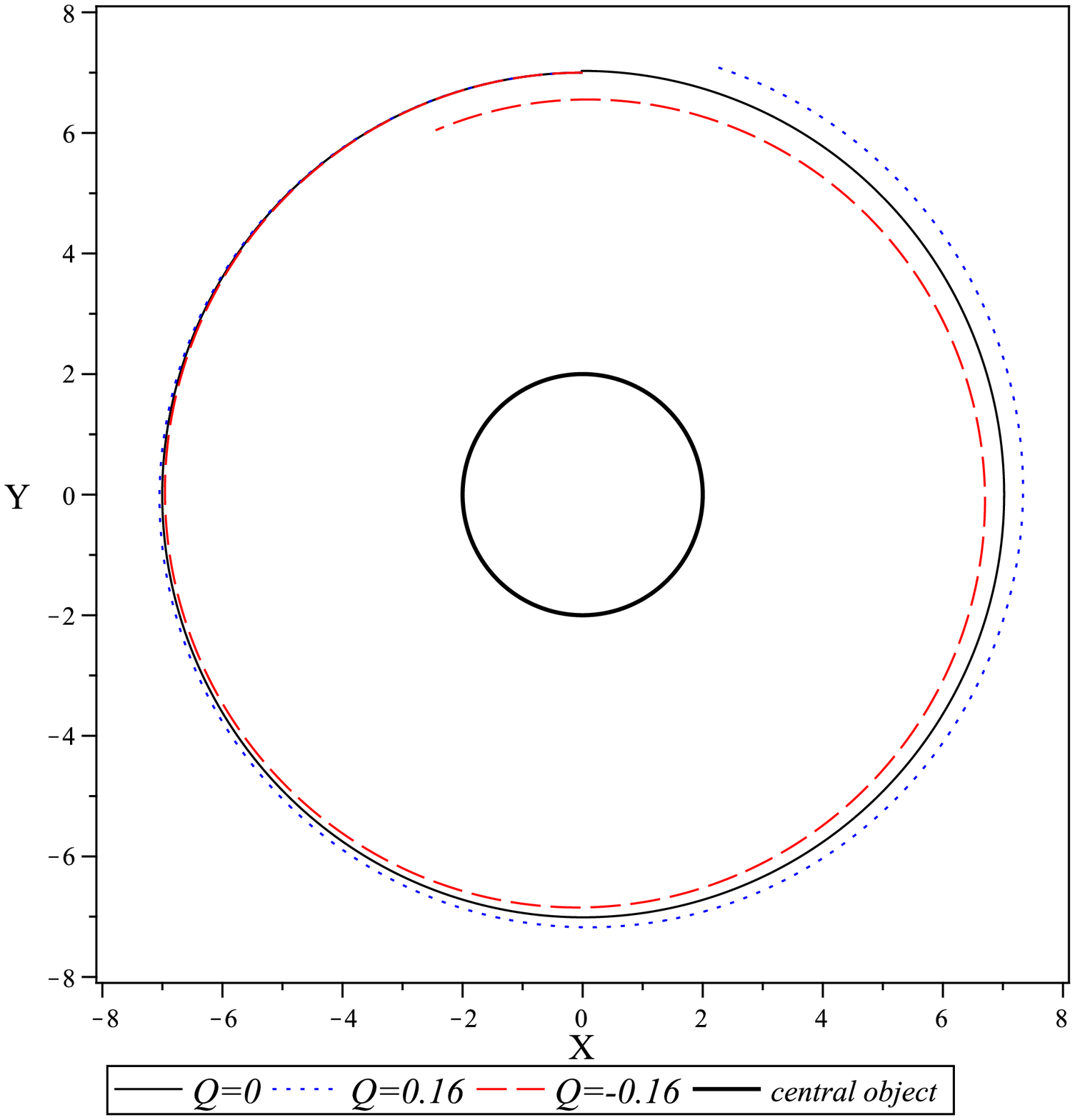}
\\[0cm] \mbox{(a)} & \mbox{(b)}\\
\includegraphics[scale=0.31]{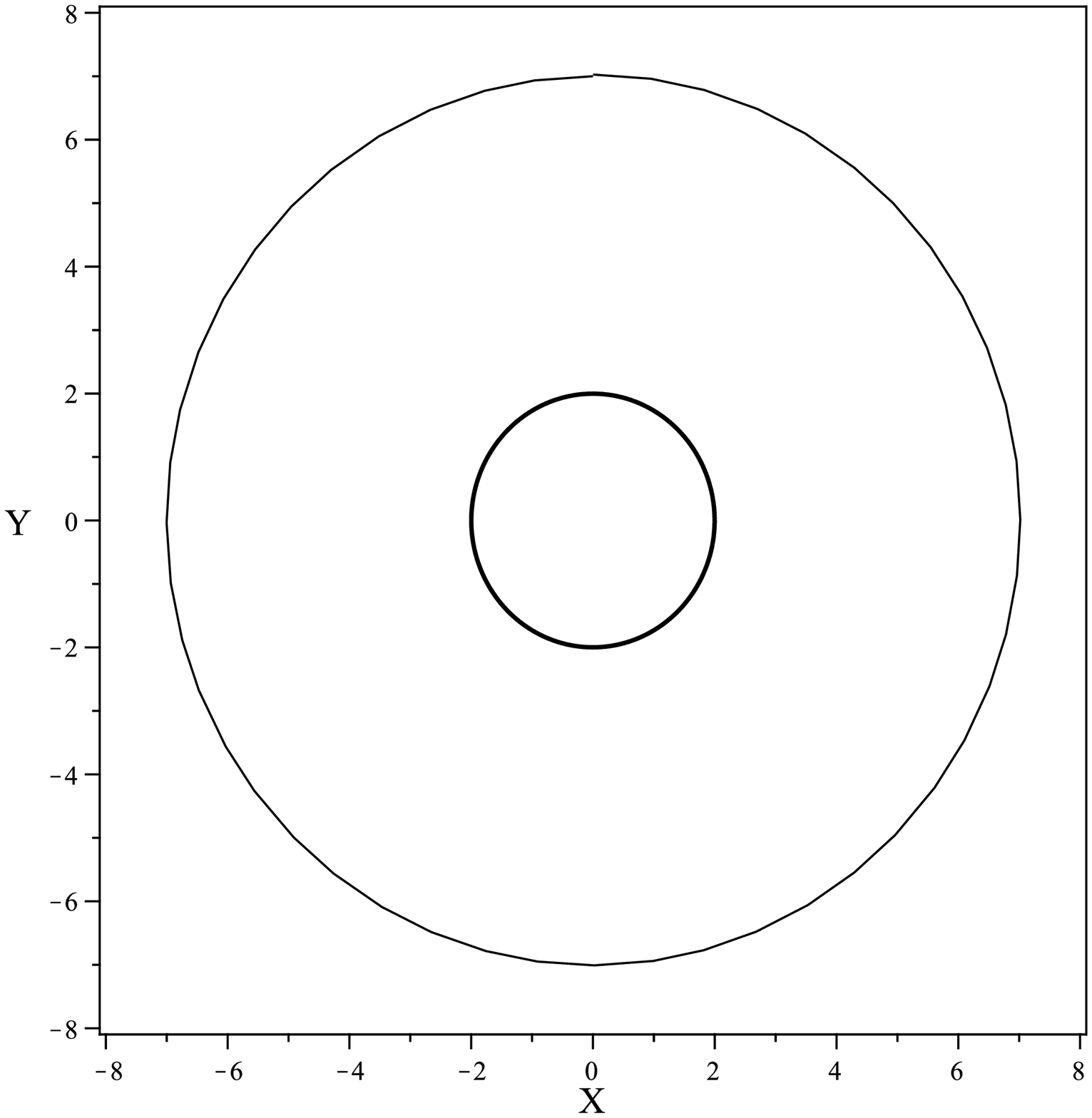}& \includegraphics[scale=0.325]{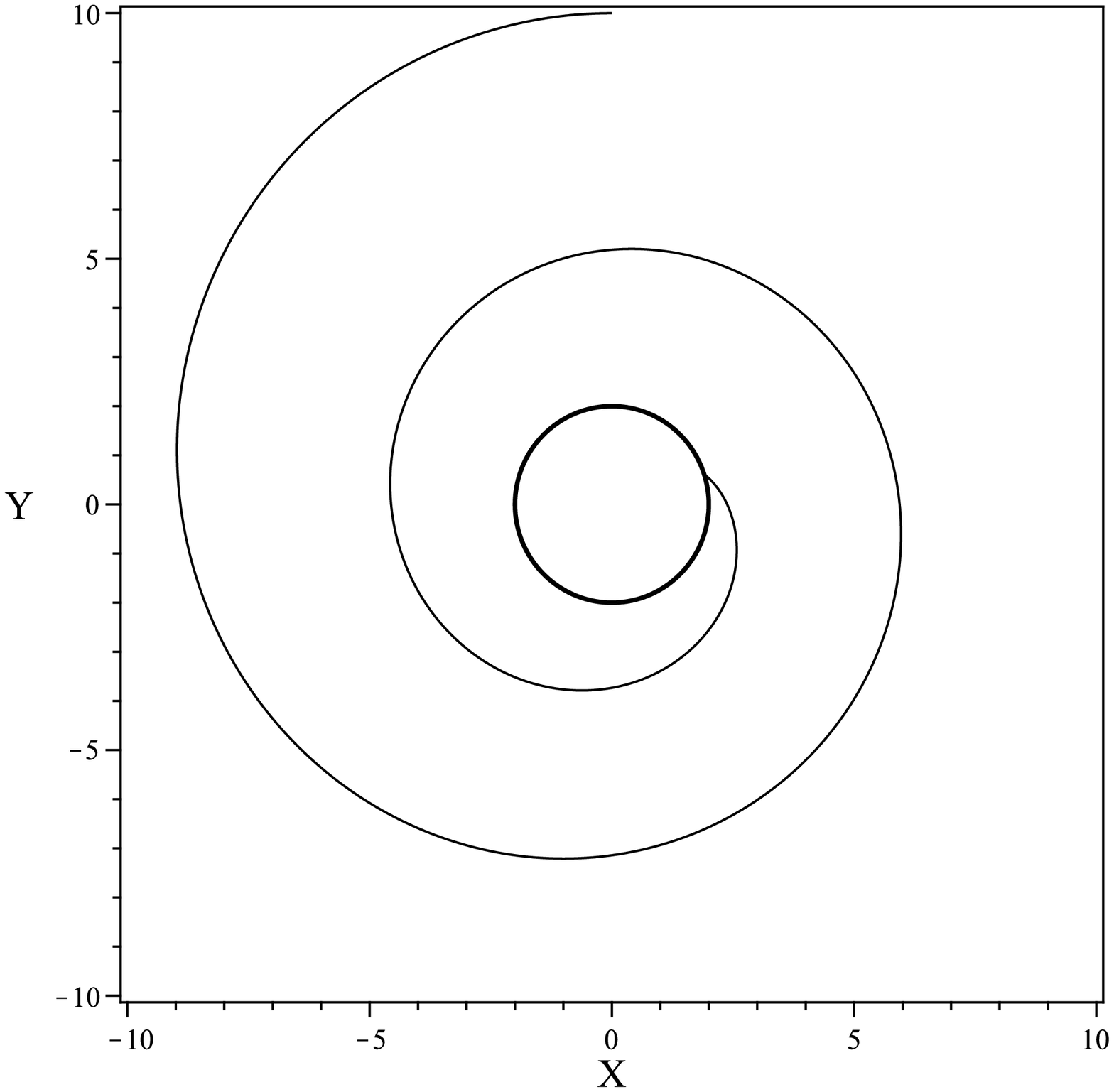}
\\[0cm]  \mbox{(c)}& \mbox{(d)}\\
\includegraphics[scale=0.325]{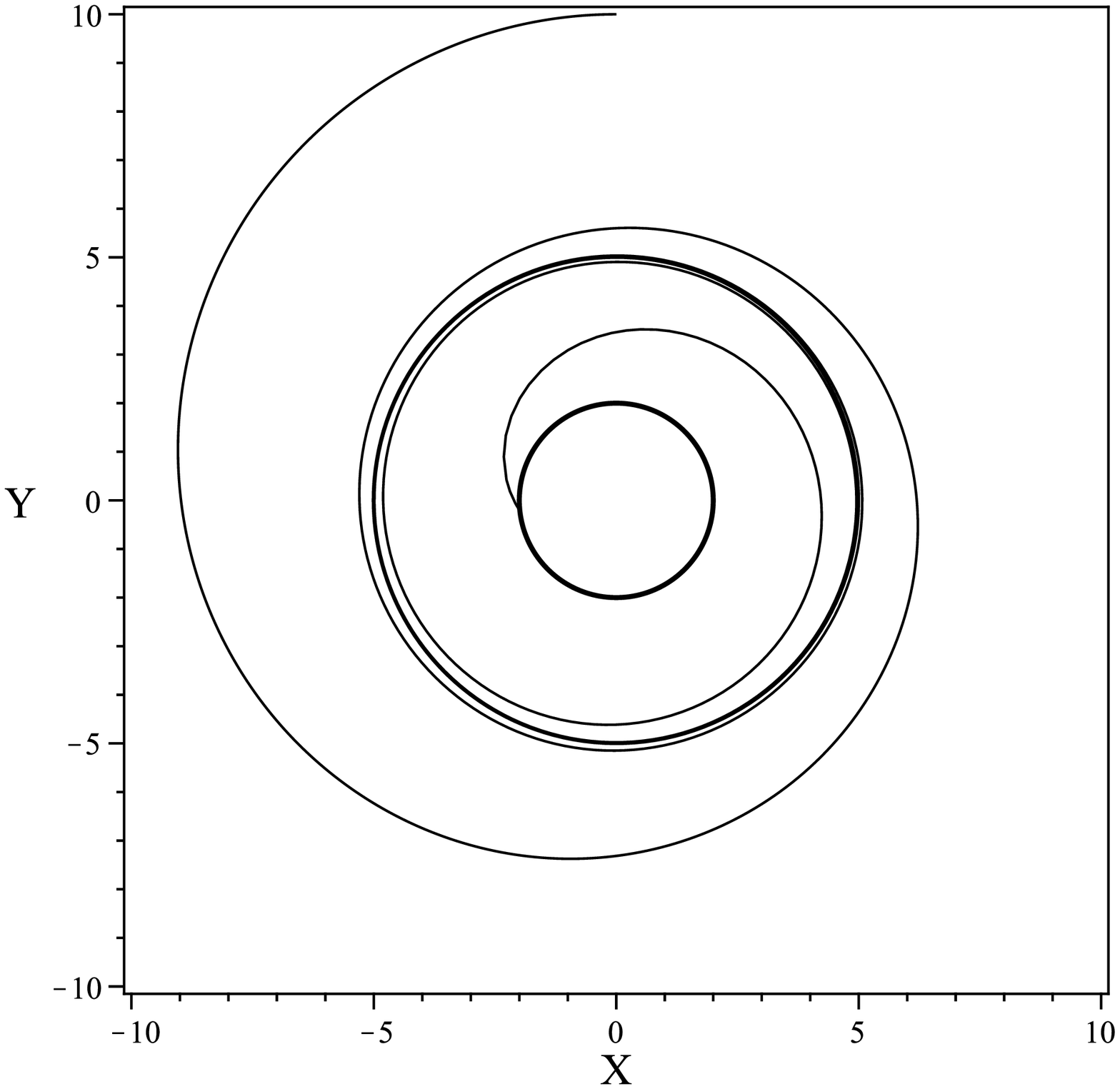}&\includegraphics[scale=0.325]{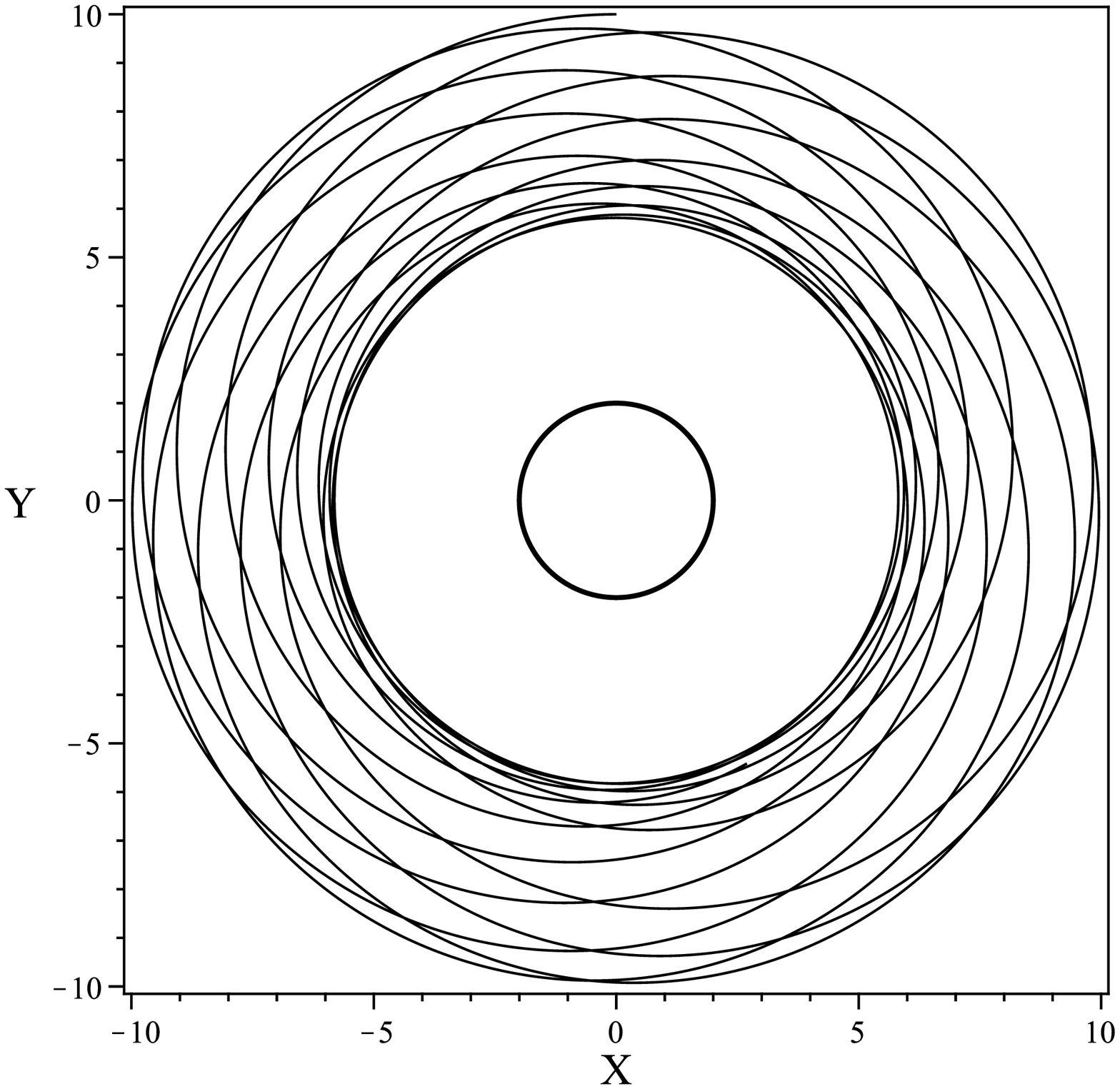}
\\[0cm] \mbox{(e)} & \mbox{(f)}
\end{array}$
{\caption{(a) One revolution of the test particle in the field of a rotating central body for $Q=0$. (b) The motion of the test particle in the field  of non-rotating deformed object for $J=0$. (c) Circular orbit with $J=-0.05,\ Q=-0.1575$. The other parameters for Figs (a), (b) and (c) are $r=7,\ d\phi/ds=0.07145$. (d) Spiral orbit for $Q=-0.16$. (e) Spiral orbit for $Q=0$. (f) Bound orbit for $Q=0.16$. For Figs. (d), (e) and (f) the other parameters are $\ J=0,\ r=10,\ d\phi/ds=0.035355$. For all Figs. these parameters are common $M=1,\ \phi=\pi/2,\ dr/ds=0$.}
\vspace{-0.7cm} 
\label{plotq}}
\end{figure}

In Fig(a) we show the differences between the geodesics with the same initial conditions arising due to the rotation of the central body i.e. the frame dragging effect in the strong field regime. The solid line for $J=0$ corresponds to equatorial circular geodesics in the Schwarzschild spacetime. The dashed line for $J>0$ corresponds to co-rotating and the dotted line for $J<0$ corresponds to contra-rotating orbits. In Fig(b) we show the differences between the geodesics with the same initial conditions arising due to the deformation of the source i.e. the oblateness of the central body. The solid line for $Q=0$ corresponds to equatorial circular geodesics in the Schwarzschild spacetime. The dashed line for $Q<0$ corresponds to the geodesics in the field of oblate and the dotted line for $Q>0$ corresponds to the geodesics in the field of the prolate central body. It is easy to see that varying the quadrupole parameter $Q$ one can recover the deviations from the Schwarzschild spacetime geodesics analogous to those caused by the frame dragging effect. By selecting the values of $J$ and $Q$ one can recover the circular orbits as in Fig. (c). In Figs (d), (e) and (f) we consider the geodesics with the same initial conditions in the field of non-rotating bodies with the increasing values of $Q$. As a result, we obtain different spiraling and bound trajectories of the test particle. For details see Fig. \ref{plotq}.    
\subsection{Radius of marginally stable, marginally bound and photon orbit} The condition $\varepsilon=-U_{t}=1$ gives the radius of the marginally bound orbit $r_{\rm mb}$, where $\varepsilon$ is the conserved specific energy per unit mass of the particle and the normalization condition  $P^{\alpha}P_{\alpha}=0$ gives the photon orbit radius $r_{\rm ph}$, where $P=\Gamma_{\rm ph}[\partial_{t}+\zeta_{\rm ph}\partial_{\phi}]$ is the photon 4-momentum. Note, that the normalization condition $P_{\alpha}P^{\alpha}=0$ gives the orbital angular velocity for the photon $\zeta_{\rm ph}$, however $\Gamma_{\rm ph}$ remains arbitrary. In order to define the photon orbit radius $r_{\rm ph}$, first, one has to define $\zeta_{\rm ph}$ and evaluate the expression for the 4-acceleration $a^{\alpha}$. For the circular geodesic  the condition $a^{\alpha}=0$ is enough to find $r_{\rm ph}$. In addition, by setting $dl/dr=0$ one can find the radius of the marginally stable orbit $r_{\rm ms}$, where $l=-U_{\phi}/U_{t}$ is the specific angular momentum per unit energy of the particle. 
\begin{eqnarray}
&&r_{\rm mb}=4M\left[1\mp\frac{1}{2}j+\left(\frac{8033}{256}-45\ln2\right)j^2+\left(-\frac{1005}{32}+45\ln2\right)q\right],\nonumber\\
&&r_{\rm ph}=3M\left[1\pm\frac{2\sqrt{3}}{9}j+\left(\frac{1751}{324}-\frac{75}{16}\ln3\right)j^2+\left(-\frac{65}{12}+\frac{75}{16}\ln3\right)q\right],\nonumber\\
&&r_{\rm ms}=6M\left[1\pm\frac{2}{3}\sqrt{\frac{2}{3}}j+\left(-\frac{251903}{2592}+240\ln\frac{3}{2}\right)j^2+\left(\frac{9325}{96}-240\ln\frac{3}{2}\right)q\right].\nonumber
\end{eqnarray}
It is clear that the presence of both the rotation and quadrupole parameters can increase or decrease the values for $r_{\rm mb},\ r_{\rm ph} $ and $r_{\rm ms}$. For the sake of comparison, if one writes these radii in the Boyer-Lindquist coordinates using the reverse of (\ref{tr2}) for $\theta=\pi/2$, and the relation (\ref{tr1}), then it is easy to obtain the following expressions for the Kerr solution with accuracy up to second order terms in the rotation parameter $a$:
\begin{equation}\label{rrr}
\vspace{-0.8cm} 
R_{\rm mb}=4M\left[1\pm\frac{a}{2M}-\frac{a^2}{16M^2}\right],\quad R_{\rm ph}=3M\left[1\mp\frac{2\sqrt{3}}{9}\frac{a}{M}-\frac{2a^2}{27M^2}\right],\nonumber\\
\end{equation}
\begin{equation}
R_{\rm ms}=6M\left[1\mp\frac{2}{3}\sqrt{\frac{2}{3}}\frac{a}{M}-\frac{7a^2}{108M^2}\right].\nonumber
\end{equation}
These radii are exactly those radii, expanded in terms up to second order in $a$, given in the work of Bardeen et. al. \cite{Bardeen1972} for the Kerr solution.
\section{Conclusion}
In this work we have explored the domain of validity of the Hartle-Thorne solution as well as the geodesics in this spacetime.
We considered equatorial circular geodesics and investigated the role of the quadrupole parameter in the motion of a test particle. Besides, we have shown that the effects arisen from the rotation of the source can be balanced (increased or decreased) by its oblateness. It would be also interesting to investigate the tidal effects in the Hartle-Thorne spacetime. This task will be treated in a future work.

\end{document}